\documentclass[submission,copyright,creativecommons]{eptcs}
\providecommand{\event}{AREA2020} 
\usepackage{graphicx}
\usepackage{underscore}           
\usepackage{times}
\usepackage{latexsym}
\usepackage{stmaryrd}
\usepackage{tikz}
\usetikzlibrary{arrows,automata,positioning}
\usepackage[all]{xy}
\usepackage{xspace}
\usepackage{xurl}
\usepackage{listings}
\usepackage{fancyvrb}

\newcommand{\atomicInt}[3]{{#1}\transmsg{#2}{#3}}
\newcommand{\prefixop}{{:}}

\newcommand{\transmsg}[1]{\stackrel{{#1}}{\Longrightarrow}}

\title{Engineering Reliable Interactions\\
in the Reality-Artificiality Continuum}
\author{Davide Ancona, Chiara Bassano, Manuela Chessa, Viviana Mascardi, Fabio Solari
\institute{University of Genova, Italy}
\institute{DIBRIS, Department of Computer Science and Technology, Bioengineering, Robotics and Systems Engineering}
\email{name.surname@unige.it}
}
\def\titlerunning{Engineering Reliable Interactions 
in the Reality-Artificiality Continuum}
\def\authorrunning{D. Ancona, C. Bassano, M. Chessa, V. Mascardi  \& F. Solari}
\begin{document}
\maketitle

\begin{abstract}
Milgram's reality-virtuality continuum applies to interaction in the physical space dimension, going from real to virtual. However, interaction has a social dimension as well, that can go from real to artificial depending on the companion with whom the user interacts. In this paper we present our vision of the Reality-Artificiality bidimensional Continuum (RAC), we identify some challenges in its design and development and we discuss how reliable interactions might be supported inside RAC. 
\end{abstract}

\section{Introduction}
According to the Merriam-Webster Dictionary, interaction is the ``mutual or reciprocal action or influence'' \cite{interaction}; its definition covers both social interaction carried out by humans and socially-capable entities via communication, and physical interaction among physical entities and/or humans. 

\begin{figure}[!ht]
    \centering
    \includegraphics[width=.7\linewidth]{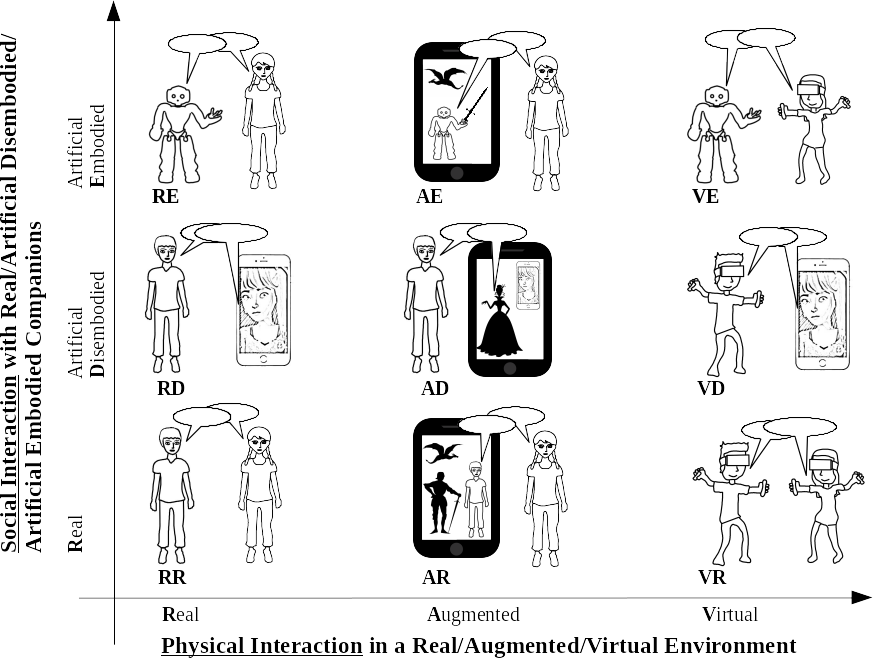}
\caption{RAC: the bidimensional Reality-Artificiality Continuum.}
    \label{fig:continuum}
\end{figure}

The reality-virtuality continuum was introduced by Milgram in 1994 \cite{Milgram94augmentedreality} and applies to interaction in the physical space dimension, going from real to virtual.   
Given that the word {\em interaction} is also strongly characterized by a social dimension, 
%
the reality-virtuality continuum paradigm should be considered inside a bidimensional space, as shown in Figure \ref{fig:continuum}, rather than inside a unidimensional one.
%
The physical dimension, drawn on the horizontal axis, involves interaction with an environment, be it real, augmented or virtual.
The social dimension, associated with the vertical axis, involves communicative  interaction via natural language with companions, be them human or artificial.

A virtual environment is composed of virtual elements, only; an augmented reality environment is the combination of both virtual and real objects. Though the virtual elements should be consistent with the real world surrounding the user for both augmented and virtual scenes, the perceptual issues and the alignment constraints are different. Indeed, in immersive virtual reality the environment is coherent since the whole environment is virtual and only an interaction coherence is required. Whereas in augmented environments the visual alignment issues can rise among virtual and real parts of the scene, even for small spatial errors.  
While Milgram's reality-virtuality continuum in the physical dimension is commonly accepted, we are not aware of an analogous ``social interaction reality-artificiality continuum'' paradigm, as depicted on the vertical axis. 
In our vision, social interaction goes from real to artificial, the latter involving humans and  artificial companions, further classified as disembodied, like chatbots and software agents, and embodied, like robots and other autonomous systems equipped with a hardware structure.  The various forms of social interaction can be combined with the various forms of physical interaction, leading to the identification of different areas in Figure \ref{fig:continuum}.  Although the best interaction style for most people is in the real physical space, and with real persons (the {\bf RR} bottom-left area in Figure \ref{fig:continuum}), sometimes this interaction is not possible. 
Before the Covid-19 pandemia, the impossibility to interact in the {\bf RR} area was mainly associated with persons living in rural and alpine areas not well connected with villages nearby, and with hospitalized or disabled people, namely with fragile categories at risk of social exclusion. Covid-19 made instead clear that billions of humans can be suddenly prevented from moving in the {\bf RR} area for a long time. Today, frailty of interactions is a global condition, and our well-being is dramatically affected by their lack: social distancing prevents interaction and the effects of prolonged isolation may worsen or trigger mental health problems\footnote{\url{https://www.sciencenews.org/article/coronavirus-covid-19-social-distancing-psychological-fallout},\\ \url{https://www.nationalgeographic.com/history/2020/04/psychologists-watching-coronavirus-social-distancing-coping/},\\ \url{https://www.forbes.com/sites/traversmark/2020/04/17/social-distancing-is-harming-mental-health-more-than-physical-health-according-to-new-survey-data/}.}.

In our futuristic vision, interaction will be possible even when no real environment, and no real companions, are available: the real or artificial companion may be immersed in a real,  augmented, or virtual environment, and the human user may interact with it in natural language. Virtual Reality (VR) and Augmented Reality (AR) may be the media through which users interact with their companions, to overcome isolation when the {\bf RR} area is not accessible.  
To make our vision feasible, the results already achieved in integrating robotics \& VR/AR should be merged with the results in the chatbots \& VR/AR field, and should be further extended to cope with the peculiar issues of both areas and with the challenges raised by their merge. We name the logical vision resulting from this merge, that we depict in Figure \ref{fig:continuum}, RAC, for Reality-Artificiality Continuum.

In this position paper, we overview some of the challenges raised by the design and development of the RAC vision. In particular, we discuss some techniques suitable to design and implement reliable interactions  in RAC, with special attention to applications related with well-being. 
We borrow the definition of reliability by \cite{DBLP:journals/corr/abs-2001-09124}: ``\emph{
For systems offering a service or function, reliability means that the service or function is available when needed. A software system is reliable to the extent that it meets its requirements  consistently, namely that it makes good decisions in all situations.}
Among the requirements we envisage for a system implementing the RAC vision, we mention

\indent 1. to be available to anyone who needs interaction, anywhere, hence, to be  based on free/open-source software and on low cost technologies; while it is clearly impossible to provide all RAC users with embodied companions, most of  them should at least be able to interact with disembodied companions in VR/AR settings;

\indent 2. to be safe, hence, to meet privacy issues, to avoid dangerous behaviors, to avoind hurting the users' feelings by always respecting their sensibility, as represented by their user's profile;

\indent 3. to be realistic and engaging, in order to provide a believable and rewarding  experience to the user, hence, to ensure the coherence between the real environment where the user in immersed and the augmented/virtual one built on top of the real one;

\indent 4. to comply with the most recent guidelines for ethics in robotics \cite{BS8611,IEEE-P7007-standard} and in autonomous systems \cite{EthicallyAlignedDesign2019,DBLP:journals/pieee/WinfieldMPE19}.

The above requirements may be further specialized for the different areas of the bidimensional RAC space. From one hand the reliability of social interactions can be preserved in immersive VR since all the agents are coherent among them since all are virtual. Whereas in augmented environments users can feel differences among virtual and real agents. On the other hand, immersive VR may not be well accepted by some people with special needs, since VR creates a completely alternative reality.

Despite these differences between the different types of interaction, which create further specific challenges, the four requirements above are the most important for our RAC vision, and should always be met. Designing and implementing a system providing (some of) the interaction functionalities sketched in Figure \ref{fig:continuum} in such a way that it meets the four requirements above is what we mean by ``engineering reliable interactions in the Reality-Artificiality Continuum''.

\section{State of the art}
\label{section:stateoftheart}
In the last 10 years, the interest of the research community towards social inclusion in general, and inclusion of fragile people in particular, has been growing year after year.  Social inclusion was one of the eleven priorities for Cohesion Policy in 2014-2020 (``thematic objective 9''\footnote{\texttt{https://ec.europa.eu/regional\_policy/en/policy/how/priorities}.}) and several European Community research programs have funded, or are going to fund in the near future, projects aiming to address these needs\footnote{\url{http://www.europeannetforinclusion.org/},\\ \url{http://www.interreg-alcotra.eu/it/decouvrir-alcotra/les-projets-finances/pro-sol},\\ \texttt{https://eacea.ec.europa.eu/erasmus-plus/news/social-inclusion-and-common-values-the-}\\ \texttt{contribution-in-the-field-of-education-and-training-2019\_en}.}. At the end of April 2020,  a third of the world population was on some form of a coronavirus lockdown, meaning their movements are restricted and controlled by their respective governments\footnote{\url{https://www.statista.com/chart/21240/enforced-covid-19-lockdowns-by-people-affected-per-country/}}: many of us became fragile, and the need for social inclusion is now affecting at least 2.5 billions people. One way to achieve social inclusion is by boosting interaction. 

The starting point of our vision is hence the Milgram's reality-virtuality continuum \cite{Milgram94augmentedreality}, or Mixed Reality (MR) continuum, a commonly accepted paradigm, which describes Virtual and Augmented Reality (VR/AR) applications like a unitary approach, rather than distinct concepts, and which is depicted in the {\bf RR}, {\bf AR}, {\bf VR} areas in Figure \ref{fig:continuum}. The MR continuum is a scale between completely real environments with no computer enhancements, towards virtual environments that are totally generated by the computer. In between these two extremes are Augmented Reality and Augmented Virtuality (AV), which are a combination of real and virtual worlds.
Similarly, the devices could go from non-immersive and standard visualization devices, e.g. monitors or tablets, to highly immersive systems, like the VR head-mounted displays, typically used for VR applications, but also for wearable AR \cite{DBLP:conf/ismar/KruijffSF10}. 

So far, the most critical issues when engineering VR/AR applications involve their perceptual issues, undesired effects, and effective use in contexts different from the entertainment: these are open and largely discussed problem, e.g. addressing the wrong estimation of distances \cite{hucapp19, 10.1145/3359996.3364245, Klinghammer2016AllocentricII, 10.1145/3009939.3009947,  maiello2015effectiveness, pointon2018affordances}. This negatively affects interaction, for example in industrial or medical training tasks, but how to interact in a natural way in such environments is also important for their acceptance, especially by fragile people \cite{bassano2018studying}. 
Indeed, immersion, presence and interactivity are all factors that influence VR acceptance and use \cite{chessa2019perceptual,DBLP:conf/amcis/MutterleinH17}. Some studies analyze acceptance of VR for specific classes of subjects, e.g. for cognitive deficits caused by diverse disorders or traumatic brain injury \cite{DBLP:journals/cmpb/CostaC04}, with positive results both in terms of patients' involvement and positive effects on their rehabilitation, also considering immersive VR head-mounted displays, or for e-Health systems for the elderly \cite{DBLP:journals/cbsn/BotellaECBGQRL09,chessa2018human}. The problem of acceptance of VR/AR has recently been studied also in education \cite{DBLP:journals/jcsci/DalimKKSB17} and tourism \cite{leue2014theoretical}. 

Moving towards the acceptance of disembodied artificial companions in the real physical environment ({\bf RD} area in Figure \ref{fig:continuum}), despite the longstanding history of studies on the connections between sentiments, emotions and interaction in natural language with conversational agents and chatbots, with ad-hoc task also in the most recent SemEval competition \cite{DBLP:conf/semeval/TafreshiD19}, studies dealing with adoption and acceptance of chatbots by humans are very recent \cite{rietz2019impact}, especially when chatbots are used in the healthcare domain  \cite{DBLP:conf/ecis/LaumerMG19}. One requirement for a conversational agent to be accepted by its user is its capability to adapt to her/him. The literature on conversation adaptation is also very recent and almost limited \cite{DBLP:conf/conversations/HeyselaarB19,DBLP:conf/insci/KlopfensteinDR18}.
%
Apart from Amazon Sumerian\footnote{\url{https://aws.amazon.com/sumerian/}.}, it seems that no frameworks for integrating chatbots into VR/AR applications ({\bf AD} and {\bf VD} areas in Figure \ref{fig:continuum}) exist, which makes the RAC vision extremely novel and original. Sumerian was launched in 2018 \cite{DBLP:conf/amcis/Fiedler19} and some authors criticize it for its lack of transparency and its questionable data policy, concluding that it is inappropriate for health-related applications \cite{8864580}. Moreover, the use of applications developed through Amazon Sumerian implies a payment, which contrasts with the first RAC requirement. 

As far as the {\bf AE} and {\bf VE} areas in Figure \ref{fig:continuum} are concerned, many solid proposals to integrate robotics and VR/AR exist, but these proposals do not take human-robot natural language interaction into account. Seminal works on robotics and VR/AR integration date back to the turn of the millennium \cite{burdea1999invited,freund1999projective} and recent scientific proposals, some of which near-market, span many different domains. 
Training is among the most lively ones: in September 2019, Guerin and  Hager patented a system and methods to create an immersive virtual environment using a VR system that receives parameters corresponding to a real-world robot that may be simulated to create a virtual robot based on the received parameters. Users may train the virtual robot, and the real-world robot may be programmed based on the virtual robot training\footnote{\url{https://patentimages.storage.googleapis.com/e4/81/62/ca61f9719139a0/US20190283248A1.pdf}.}. Just one month later the Toyota Research Institute released a video showing how they use VR to allow human trainers to teach robots in-home tasks\footnote{\url{https://youtu.be/6IGCIjp2bn4}.}.
The just published paper by Makhataeva and Varol \cite{arr2020} surveys about 100 AR applications in medical robotics, robot control and planning, human-robot interaction and robot swarms from 2015 and 2019. Recent experiments demonstrate the value in using modern VR/AR technologies to mediate human-robot interactions \cite{10.1007} and boost robots acceptance by their users.

Finally, we mention the very active field of natural language human-robot interaction, positioned in the  top-left {\bf RE} area in Figure \ref{fig:continuum}. 
Robots must {\em understand} the natural language, and robotic systems  capable  of  executing  complex  commands  issued  in  natural language  that  ground  into  robotic  percepts have been designed and implemented \cite{guadarrama2013grounding},  grammar based approaches to human-robot interaction have been experimented and compared with data driven ones \cite{bastianelli2014effective}, algorithms targetted to recognize motion verbs such as ``follow'', ``meet'', ``avoid'' have been proposed \cite{kollar2014grounding}.
However, to be believable and pleasant companions, robots should also {\em generate} natural language. The  reasons why so few current social robots make use of  sophisticated generation techniques are discussed by Foster in her paper on ``natural language generation for social robotics: opportunities and challenges'' \cite{foster2019natural}.

\section{Engineering Reliable Interactions in RAC}
\label{section:vision}

The analysis of the state of the art shows that many building blocks that are needed to implement the vision put forward in Figure \ref{fig:continuum} are already available to the research community, at least as working prototypes. Nevertheless, the RAC building is still to come. 

We believe that many years will pass by before we will see antropomorphic robots interacting with humans in natural language inside a virtual reality ({\bf VE} area in Figure \ref{fig:continuum}, the most challenging one). However, some approaches may boost the integrated RAC vision take off, and allow its developers to engineer interactions that are reliable. In this section we review some of them.

The different combinations of physical and social interaction raise different research challenges and offer different opportunities for the users. As an example, a pure virtual experience could be well accepted by a young user,  but it might not be applied to elderly, who might more benefit from a communication-intensive, voice-based interaction where words are carefully selected from a vocabulary of words that are familiar to that specific person, and that (s)he can easily recognize and understand; for this kind of person, interaction with the augmented reality should not be an immersive one.

To engineer reliable interactions in RAC, users should be profiled, the real environment where they operate should be recognized and abstracted into a model, and interactions should dynamically adapt to their profile and real environment, which may change over time, and at the same time should follow some patterns, or protocols, that are ``safe by design''.

We disregard the profiling issue, 
already well covered by the literature \cite{eke2019survey,rossi2017user}, and we discuss how reliability of interaction could be addressed by exploiting ontologies and runtime verification.\\[-.15cm]


\noindent {\bf Making virtual and real environments semantically consistent via ontologies.}
In order to make interactions believable and suitable to the human user, even when taking place in the most challenging {\bf VE} area, the  immersive virtual environment should be consistent with the real world surrounding the user. 
We believe that ontologies may help in achieving this goal.
``{\em In the context of computer and information sciences, an ontology defines a set of representational primitives with which to model a domain of knowledge or discourse.  The representational primitives are typically classes (or sets), attributes (or properties), and relationships (or relations among class members). } ''
\cite{gruber2008ontologies}.

Once the objects in the real environment are detected and the corresponding bounding box created, the virtual scenes should be composed on-the-fly\footnote{The bounding box is the volume that fully encloses the real object. A virtual scene is a digital counterpart of a real scene, where the objects are virtual and not real.}, and in a (semi-)automatic way, by inserting objects of the same shape, position and spatial properties of the corresponding real ones. These virtual objects should be  semantically consistent with the virtual scene.
The scene is composed on-the-fly in order to handle objects that can move in the real scene and it is necessary an automatic procedure to carry out this process. A human operator can contribute to the creation of a virtual scene by working on the semantic consistency of the objects and on the kind of environment that the operator wants to implement.  The whole scene can be built off-line to decrease the computational cost and to handle on-the-fly only the objects that can change their position in the real scene.  

For example, if the virtual scene represents a bush, and a real object has been detected with a parallelepipedal bounding box starting from the ground and 70cm tall, it might be substituted, in the virtual scene, by a shrub. If the height is 1.70cm, it might be substituted by a tree.
If the virtual scene represents a shore, those two objects might be substituted by a chair and a beach umbrella, respectively, and if it is a urban environment, they might be a trash bin and a pole with a road sign.

In order to associate virtual objects with real objects in a semantically consistent way, we need to {\em represent the knowledge about which virtual objects make sense into a virtual scene}, also depending on their geometric features, via one or more ontologies. To the best of our knowledge, the connections between VR/AR and ontologies have been explored in a very limited way, and the idea of exploiting ontologies in VR/AR to make virtual and real objects consistent is an original one. The most similar proposal dates back to 2011 and applies to 3D virtual worlds; it aims at labelling some things with tags like ``chair'' and  ``table'' and associate functional specifications to enable computers to reason about them, but no real-to-virtual semantic alignment is foreseen \cite{DBLP:journals/internet/EnoT11}.

Once ontologies enter the RAC, they can also serve as a means to ensure semantic interoperability among different users and to ease natural language interaction. This would be a more standard way to use ontologies in RAC, since the relationships between natural language processing and ontologies have been studied since the advent of the semantic web \cite{maynard2016natural,mellish2005semantic,wilks2009natural}.

The exploitation of ontologies makes both physical and social interaction reliable because, as observed by Uschold and Gruninger more than 20 years ago,  ``{\em a formal representation makes possible the automation
of consistency checking resulting in more reliable software}'' \cite{uschold1996ontologies}.

As an example, by writing disjunction axioms in the ontology it would be possible to formally ensure that a physical object is never rendered with a virtual appearance, if that virtual look may hurt a user with some given profile. 
In a similar way, ontologies might ensure that the words selected by the artificial companion when generating sentences are compliant with the profile of the user, and that they are properly disambiguated when uttered by the user. To make an example, a user with a young profile might use ``paw'' as a slang acronym for ``Parents Are Watching'' while a cat lady would use it with its most common meaning. 
By avoiding misunderstandings and choices of words or virtual shapes unsuitable for some categories of users, reliability of interactions would increase.   \\[-.15cm]

\noindent {\bf Verifying social and physical interactions at runtime.}
The other approach that we envision as a way to make interactions in RAC more reliable,  is runtime verification.

Runtime verification is a formal method aimed at checking the
current run of the system. It takes as input a data stream of time-stamped
sensor or software values and a requirement to verify, expressed in the form of a temporal logic formula, most commonly Linear Time Temporal Logic \cite{DBLP:journals/tosem/BauerLS11}, Mission-time Linear Temporal
Logic (MLTL) \cite{LVR19},  First-Order Linear Temporal Logic \cite{BKMP08}, or via  trace-based formalisms like Trace Expressions \cite{DBLP:conf/birthday/AnconaFM16}; the output is a stream of $\left< \mathrm{time}, \mathrm{verdict} \right>$ tuples containing a time stamp and the verdict from the valuation of the requirement (true, 
false, or unknown when a three valued logic is adopted), evaluated over the data stream starting from that time step.

The runtime verification formalism and engine we have been developing in the last seven years, Trace Expressions, has proven useful both to monitor agent interaction protocols (AIP, \cite{el1999protocol,odell2000representing}) taking place in purely software multiagent systems \cite{DBLP:conf/atal/AnconaFM17,DBLP:conf/atal/FerrandoAM17}, also when self adaptation of agents is a requirement \cite{DBLP:conf/atal/AnconaBFM15},
and to verify event patterns taking place in physical environments like the Internet of Things \cite{DBLP:conf/icwe/LeottaAFORR18,DBLP:conf/enase/LeottaCFOARR19}. The Runtime Monitoring Language, RML\footnote{\url{https://rmlatdibris.github.io/}}, is a synthesis of these years of research: it decouples monitoring from instrumentation by allowing users to write specifications and to synthesize monitors from them, independently of the system under scrutiny and its instrumentation.

RML would be suitable to monitor physical interactions in any portion of the RAC bidimentional space. To make an example inspired by the current lockdown experience, the number of human users in the same room at the same time should be no more than 2, they should keep a 1m distance, and no user could stay in the room for more than 45 minutes. If users were equipped with sensors suitable for indoor localization, RML could monitor the balance of ``user X enters the room at time T1'' and ``user X leaves the room at time T2'' events and check that the number of users in the room and the time spent there meet the lockdown rules. A further continuous check that if two users are in the room, they keep the mandatory distance, could also be easily implemented. 

What would be definitely harder, but worth exploring, is the runtime monitoring of natural language interactions. Let us consider the following situation: Alice is interacting with her artificial companion, which is aware that Alice profile is ``very polite''. The artificial companion expects that after a greeting, Alice answers with a greeting, and after the accomplishment of some task, Alice thanks. There are many ways Alice can greet her companion or express her gratitude, so there are many different concrete utterances that should be tagged ad ``greeting'' or ``gratitude''. It would be up to some sophisticated natural language processing component to tag utterances in the right way. Assuming that utterances are managed as events, each tagged with its event type ($greets$, $thanks$), the interaction protocol with very polite users might require that 
$\atomicInt{robot}{greets}{alice}~ \prefixop~ \atomicInt{alice}{greets}{robot}~\prefixop~\epsilon
$
and 
$
\atomicInt{robot}{task\_completed}{alice}~\prefixop~ \atomicInt{alice}{thanks}{robot}~\prefixop~\epsilon
$ 
where $\atomicInt{A1}{Tag}{A2}$ means that $A1$ uttered a sentence tagged as  
$Tag$ to $A2$, and $\prefixop$ is the prefix operator meaning that after one event, another must take place. $\epsilon$ represents the end of the interaction. 

Being able to correctly tag utterances on the fly would allow RML to detect violations of the protocol. For example, Alice might not thank her companion for some task it carried out. While this may be due to any reason, not necessarily a serious one, it might also be a signal of something wrong in Alice' mood, and might require some special action to take, or some ad hoc sub-protocol to be triggered. 

The idea that some sentences can be collapsed into the same category, and hence tagged with the same ``event type'', is already supported by many languages for building chatbots. As an example AIML, the Artificial Intelligence Markup Language\footnote{\url{http://www.aiml.foundation/doc.html}}, provides the \texttt{srai} mechanism for this purpose, and IBM Watson Assistant\footnote{\url{https://www.ibm.com/cloud/watson-assistant/}} achieves the same goal via the notion of ``intent'' exemplified by a list of sentences. 

At the state of the art, runtime verification of natural language interactions is out of reach. However, by integrating existing approaches to chatbot development and RML we might be able to move a step forward in this direction. \\[-.15cm]


As a final remark, we observe that although the RAC integrated vision is still far to come, we believe that it represents a promising direction towards the satisfaction of anywhere-anytime social inclusion needs. Many ethical, social, psychological and medical aspects should be addressed if the RAC became available, besides the technical ones.
As part of our close future work, we are going to address some of the latter: on the one hand, we will work on the alignment of virtual and real objects via ontologies. On the other, we will explore the possibility to integrate runtime verification mechanisms in human-chatbot interactions. \\[-.15cm]

 \noindent {\bf Acknowledgements.} We gratefully acknowledge Simone Ancona for the drawings of Figure \ref{fig:continuum}.


\begin{thebibliography}{10}
\providecommand{\bibitemdeclare}[2]{}
\providecommand{\surnamestart}{}
\providecommand{\surnameend}{}
\providecommand{\urlprefix}{Available at }
\providecommand{\url}[1]{\texttt{#1}}
\providecommand{\href}[2]{\texttt{#2}}
\providecommand{\urlalt}[2]{\href{#1}{#2}}
\providecommand{\doi}[1]{doi:\urlalt{http://dx.doi.org/#1}{#1}}
\providecommand{\bibinfo}[2]{#2}

\bibitemdeclare{inproceedings}{DBLP:conf/atal/AnconaBFM15}
\bibitem{DBLP:conf/atal/AnconaBFM15}
\bibinfo{author}{Davide \surnamestart Ancona\surnameend},
  \bibinfo{author}{Daniela \surnamestart Briola\surnameend},
  \bibinfo{author}{Angelo \surnamestart Ferrando\surnameend} \&
  \bibinfo{author}{Viviana \surnamestart Mascardi\surnameend}
  (\bibinfo{year}{2015}): \emph{\bibinfo{title}{Global Protocols as First Class
  Entities for Self-Adaptive Agents}}.
\newblock In \bibinfo{editor}{Gerhard \surnamestart Weiss\surnameend},
  \bibinfo{editor}{Pinar \surnamestart Yolum\surnameend},
  \bibinfo{editor}{Rafael~H. \surnamestart Bordini\surnameend} \&
  \bibinfo{editor}{Edith \surnamestart Elkind\surnameend}, editors: {\sl
  \bibinfo{booktitle}{Proceedings of the 2015 International Conference on
  Autonomous Agents and Multiagent Systems, {AAMAS} 2015, Istanbul, Turkey, May
  4-8, 2015}}, \bibinfo{publisher}{{ACM}}, pp. \bibinfo{pages}{1019--1029}.
\newblock \urlprefix\url{http://dl.acm.org/citation.cfm?id=2773282}.

\bibitemdeclare{inproceedings}{DBLP:conf/birthday/AnconaFM16}
\bibitem{DBLP:conf/birthday/AnconaFM16}
\bibinfo{author}{Davide \surnamestart Ancona\surnameend},
  \bibinfo{author}{Angelo \surnamestart Ferrando\surnameend} \&
  \bibinfo{author}{Viviana \surnamestart Mascardi\surnameend}
  (\bibinfo{year}{2016}): \emph{\bibinfo{title}{Comparing Trace Expressions and
  Linear Temporal Logic for Runtime Verification}}.
\newblock In \bibinfo{editor}{Erika \surnamestart
  {\'{A}}brah{\'{a}}m\surnameend}, \bibinfo{editor}{Marcello~M. \surnamestart
  Bonsangue\surnameend} \& \bibinfo{editor}{Einar~Broch \surnamestart
  Johnsen\surnameend}, editors: {\sl \bibinfo{booktitle}{Theory and Practice of
  Formal Methods - Essays Dedicated to Frank de Boer on the Occasion of His
  60th Birthday}}, {\sl \bibinfo{series}{Lecture Notes in Computer Science}}
  \bibinfo{volume}{9660}, \bibinfo{publisher}{Springer}, pp.
  \bibinfo{pages}{47--64}, \doi{10.1007/978-3-319-30734-3\_6}.

\bibitemdeclare{inproceedings}{DBLP:conf/atal/AnconaFM17}
\bibitem{DBLP:conf/atal/AnconaFM17}
\bibinfo{author}{Davide \surnamestart Ancona\surnameend},
  \bibinfo{author}{Angelo \surnamestart Ferrando\surnameend} \&
  \bibinfo{author}{Viviana \surnamestart Mascardi\surnameend}
  (\bibinfo{year}{2017}): \emph{\bibinfo{title}{Parametric Runtime Verification
  of Multiagent Systems}}.
\newblock In \bibinfo{editor}{Kate \surnamestart Larson\surnameend},
  \bibinfo{editor}{Michael \surnamestart Winikoff\surnameend},
  \bibinfo{editor}{Sanmay \surnamestart Das\surnameend} \&
  \bibinfo{editor}{Edmund~H. \surnamestart Durfee\surnameend}, editors: {\sl
  \bibinfo{booktitle}{Proceedings of the 16th Conference on Autonomous Agents
  and MultiAgent Systems, {AAMAS} 2017, S{\~{a}}o Paulo, Brazil, May 8-12,
  2017}}, \bibinfo{publisher}{{ACM}}, pp. \bibinfo{pages}{1457--1459}.
\newblock \urlprefix\url{http://dl.acm.org/citation.cfm?id=3091328}.

\bibitemdeclare{inproceedings}{BKMP08}
\bibitem{BKMP08}
\bibinfo{author}{David~A. \surnamestart Basin\surnameend},
  \bibinfo{author}{Felix \surnamestart Klaedtke\surnameend},
  \bibinfo{author}{Samuel \surnamestart M{\"{u}}ller\surnameend} \&
  \bibinfo{author}{Birgit \surnamestart Pfitzmann\surnameend}
  (\bibinfo{year}{2008}): \emph{\bibinfo{title}{Runtime Monitoring of Metric
  First-order Temporal Properties}}.
\newblock In: {\sl \bibinfo{booktitle}{Proceedings of 28th {IARCS} Conference
  on Foundations of Software Technology and Theoretical Computer Science
  ({FSTTCS}'08)}}, pp. \bibinfo{pages}{49--60},
  \doi{10.4230/LIPIcs.FSTTCS.2008.1740}.

\bibitemdeclare{inproceedings}{bassano2018studying}
\bibitem{bassano2018studying}
\bibinfo{author}{Chiara \surnamestart Bassano\surnameend},
  \bibinfo{author}{Fabio \surnamestart Solari\surnameend} \&
  \bibinfo{author}{Manuela \surnamestart Chessa\surnameend}
  (\bibinfo{year}{2018}): \emph{\bibinfo{title}{Studying Natural Human-computer
  Interaction in Immersive Virtual Reality: {A} Comparison between Actions in
  the Peripersonal and in the Near-action Space}}.
\newblock In \bibinfo{editor}{Paul \surnamestart Richard\surnameend},
  \bibinfo{editor}{Manuela \surnamestart Chessa\surnameend} \&
  \bibinfo{editor}{Jos{\'{e}} \surnamestart Braz\surnameend}, editors: {\sl
  \bibinfo{booktitle}{Proceedings of the 13th International Joint Conference on
  Computer Vision, Imaging and Computer Graphics Theory and Applications
  {(VISIGRAPP} 2018) - Volume 2: HUCAPP, Funchal, Madeira, Portugal, January
  27-29, 2018}}, \bibinfo{publisher}{SciTePress}, pp.
  \bibinfo{pages}{108--115}, \doi{10.5220/0006622701080115}.

\bibitemdeclare{inproceedings}{bastianelli2014effective}
\bibitem{bastianelli2014effective}
\bibinfo{author}{Emanuele \surnamestart Bastianelli\surnameend},
  \bibinfo{author}{Giuseppe \surnamestart Castellucci\surnameend},
  \bibinfo{author}{Danilo \surnamestart Croce\surnameend},
  \bibinfo{author}{Roberto \surnamestart Basili\surnameend} \&
  \bibinfo{author}{Daniele \surnamestart Nardi\surnameend}
  (\bibinfo{year}{2014}): \emph{\bibinfo{title}{Effective and Robust Natural
  Language Understanding for Human-Robot Interaction}}.
\newblock In \bibinfo{editor}{Torsten \surnamestart Schaub\surnameend},
  \bibinfo{editor}{Gerhard \surnamestart Friedrich\surnameend} \&
  \bibinfo{editor}{Barry \surnamestart O'Sullivan\surnameend}, editors: {\sl
  \bibinfo{booktitle}{{ECAI} 2014 - 21st European Conference on Artificial
  Intelligence, 18-22 August 2014, Prague, Czech Republic - Including
  Prestigious Applications of Intelligent Systems {(PAIS} 2014)}}, {\sl
  \bibinfo{series}{Frontiers in Artificial Intelligence and Applications}}
  \bibinfo{volume}{263}, \bibinfo{publisher}{{IOS} Press}, pp.
  \bibinfo{pages}{57--62}, \doi{10.3233/978-1-61499-419-0-57}.

\bibitemdeclare{article}{DBLP:journals/tosem/BauerLS11}
\bibitem{DBLP:journals/tosem/BauerLS11}
\bibinfo{author}{Andreas \surnamestart Bauer\surnameend},
  \bibinfo{author}{Martin \surnamestart Leucker\surnameend} \&
  \bibinfo{author}{Christian \surnamestart Schallhart\surnameend}
  (\bibinfo{year}{2011}): \emph{\bibinfo{title}{Runtime Verification for {LTL}
  and {TLTL}}}.
\newblock {\sl \bibinfo{journal}{{ACM} Trans. Softw. Eng. Methodol.}}
  \bibinfo{volume}{20}(\bibinfo{number}{4}), pp. \bibinfo{pages}{14:1--14:64},
  \doi{10.1145/2000799.2000800}.

\bibitemdeclare{article}{DBLP:journals/cbsn/BotellaECBGQRL09}
\bibitem{DBLP:journals/cbsn/BotellaECBGQRL09}
\bibinfo{author}{Cristina \surnamestart Botella\surnameend},
  \bibinfo{author}{Ernestina \surnamestart Etchemendy\surnameend},
  \bibinfo{author}{Diana \surnamestart Castilla\surnameend},
  \bibinfo{author}{Rosa~Mar{\'{\i}}a \surnamestart Ba{\~{n}}os\surnameend},
  \bibinfo{author}{Azucena \surnamestart Garc{\'{\i}}a{-}Palacios\surnameend},
  \bibinfo{author}{Soledad \surnamestart Quero\surnameend},
  \bibinfo{author}{Mariano~Alca{\~{n}}iz \surnamestart Raya\surnameend} \&
  \bibinfo{author}{Jos{\'{e}}~Antonio \surnamestart Lozano\surnameend}
  (\bibinfo{year}{2009}): \emph{\bibinfo{title}{An e-Health System for the
  Elderly (Butler Project): {A} Pilot Study on Acceptance and Satisfaction}}.
\newblock {\sl \bibinfo{journal}{Cyberpsychology Behav. Soc. Netw.}}
  \bibinfo{volume}{12}(\bibinfo{number}{3}), pp. \bibinfo{pages}{255--262},
  \doi{10.1089/cpb.2008.0325}.

\bibitemdeclare{misc}{BS8611}
\bibitem{BS8611}
\bibinfo{author}{\surnamestart {British Standards Institution
  (BSI)}\surnameend} (\bibinfo{year}{2016}): \emph{\bibinfo{title}{{BS 8611} --
  Robots and Robotic Devices --- Guide to the ethical design and application}}.
\newblock
  \urlprefix\url{https://shop.bsigroup.com/ProductDetail/?pid=000000000030320089}.

\bibitemdeclare{article}{burdea1999invited}
\bibitem{burdea1999invited}
\bibinfo{author}{Grigore~C. \surnamestart Burdea\surnameend}
  (\bibinfo{year}{1999}): \emph{\bibinfo{title}{Invited review: the synergy
  between virtual reality and robotics}}.
\newblock {\sl \bibinfo{journal}{{IEEE} Trans. Robotics Autom.}}
  \bibinfo{volume}{15}(\bibinfo{number}{3}), pp. \bibinfo{pages}{400--410},
  \doi{10.1109/70.768174}.

\bibitemdeclare{conference}{hucapp19}
\bibitem{hucapp19}
\bibinfo{author}{Andrea \surnamestart Canessa\surnameend},
  \bibinfo{author}{Paolo \surnamestart Casu\surnameend}, \bibinfo{author}{Fabio
  \surnamestart Solari\surnameend} \& \bibinfo{author}{Manuela \surnamestart
  Chessa\surnameend} (\bibinfo{year}{2019}): \emph{\bibinfo{title}{Comparing
  Real Walking in Immersive Virtual Reality and in Physical World using Gait
  Analysis}}.
\newblock In: {\sl \bibinfo{booktitle}{Proceedings of the 14th International
  Joint Conference on Computer Vision, Imaging and Computer Graphics Theory and
  Applications - Volume 2: HUCAPP,}}, \bibinfo{organization}{INSTICC},
  \bibinfo{publisher}{SciTePress}, pp. \bibinfo{pages}{121--128},
  \doi{10.5220/0007380901210128}.

\bibitemdeclare{inproceedings}{chessa2018human}
\bibitem{chessa2018human}
\bibinfo{author}{Manuela \surnamestart Chessa\surnameend},
  \bibinfo{author}{Chiara \surnamestart Bassano\surnameend},
  \bibinfo{author}{Elisa \surnamestart Gusai\surnameend},
  \bibinfo{author}{Alice~E. \surnamestart Martis\surnameend} \&
  \bibinfo{author}{Fabio \surnamestart Solari\surnameend}
  (\bibinfo{year}{2018}): \emph{\bibinfo{title}{Human-Computer Interaction
  Approaches for the Assessment and the Practice of the Cognitive Capabilities
  of Elderly People}}.
\newblock In \bibinfo{editor}{Laura \surnamestart
  Leal{-}Taix{\'{e}}\surnameend} \& \bibinfo{editor}{Stefan \surnamestart
  Roth\surnameend}, editors: {\sl \bibinfo{booktitle}{Computer Vision - {ECCV}
  2018 Workshops - Munich, Germany, September 8-14, 2018, Proceedings, Part
  {VI}}}, {\sl \bibinfo{series}{Lecture Notes in Computer Science}}
  \bibinfo{volume}{11134}, \bibinfo{publisher}{Springer}, pp.
  \bibinfo{pages}{66--81}, \doi{10.1007/978-3-030-11024-6\_5}.

\bibitemdeclare{article}{chessa2019perceptual}
\bibitem{chessa2019perceptual}
\bibinfo{author}{Manuela \surnamestart Chessa\surnameend},
  \bibinfo{author}{Guido \surnamestart Maiello\surnameend},
  \bibinfo{author}{Alessia \surnamestart Borsari\surnameend} \&
  \bibinfo{author}{Peter~J. \surnamestart Bex\surnameend}
  (\bibinfo{year}{2019}): \emph{\bibinfo{title}{The Perceptual Quality of the
  Oculus Rift for Immersive Virtual Reality}}.
\newblock {\sl \bibinfo{journal}{Human-Computer Interaction}}
  \bibinfo{volume}{34}(\bibinfo{number}{1}), pp. \bibinfo{pages}{51--82},
  \doi{10.1080/07370024.2016.1243478}.

\bibitemdeclare{article}{DBLP:journals/cmpb/CostaC04}
\bibitem{DBLP:journals/cmpb/CostaC04}
\bibinfo{author}{Rosa Maria Esteves~Moreira \surnamestart da~Costa\surnameend}
  \& \bibinfo{author}{Lu{\'{\i}}s Alfredo~Vidal \surnamestart
  de~Carvalho\surnameend} (\bibinfo{year}{2004}): \emph{\bibinfo{title}{The
  acceptance of virtual reality devices for cognitive rehabilitation: a report
  of positive results with schizophrenia}}.
\newblock {\sl \bibinfo{journal}{Comput. Methods Programs Biomed.}}
  \bibinfo{volume}{73}(\bibinfo{number}{3}), pp. \bibinfo{pages}{173--182},
  \doi{10.1016/S0169-2607(03)00066-X}.

\bibitemdeclare{article}{DBLP:journals/jcsci/DalimKKSB17}
\bibitem{DBLP:journals/jcsci/DalimKKSB17}
\bibinfo{author}{Che Samihah~Che \surnamestart Dalim\surnameend},
  \bibinfo{author}{Hoshang \surnamestart Kolivand\surnameend},
  \bibinfo{author}{Huda \surnamestart Kadhim\surnameend},
  \bibinfo{author}{Mohd~Shahrizal \surnamestart Sunar\surnameend} \&
  \bibinfo{author}{Mark \surnamestart Billinghurst\surnameend}
  (\bibinfo{year}{2017}): \emph{\bibinfo{title}{Factors Influencing the
  Acceptance of Augmented Reality in Education: {A} Review of the Literature}}.
\newblock {\sl \bibinfo{journal}{J. Comput. Sci.}}
  \bibinfo{volume}{13}(\bibinfo{number}{11}), pp. \bibinfo{pages}{581--589},
  \doi{10.3844/jcssp.2017.581.589}.

\bibitemdeclare{article}{eke2019survey}
\bibitem{eke2019survey}
\bibinfo{author}{Christopher~Ifeanyi \surnamestart Eke\surnameend},
  \bibinfo{author}{Azah~Anir \surnamestart Norman\surnameend},
  \bibinfo{author}{Liyana \surnamestart Shuib\surnameend} \&
  \bibinfo{author}{Henry~Friday \surnamestart Nweke\surnameend}
  (\bibinfo{year}{2019}): \emph{\bibinfo{title}{A survey of user profiling:
  state-of-the-art, challenges, and solutions}}.
\newblock {\sl \bibinfo{journal}{IEEE Access}} \bibinfo{volume}{7}, pp.
  \bibinfo{pages}{144907--144924}, \doi{10.1109/ACCESS.2019.2944243}.

\bibitemdeclare{article}{DBLP:journals/internet/EnoT11}
\bibitem{DBLP:journals/internet/EnoT11}
\bibinfo{author}{Joshua \surnamestart Eno\surnameend} \&
  \bibinfo{author}{Craig~W. \surnamestart Thompson\surnameend}
  (\bibinfo{year}{2011}): \emph{\bibinfo{title}{Virtual and Real-World Ontology
  Services}}.
\newblock {\sl \bibinfo{journal}{{IEEE} Internet Comput.}}
  \bibinfo{volume}{15}(\bibinfo{number}{5}), pp. \bibinfo{pages}{46--52},
  \doi{10.1109/MIC.2011.75}.

\bibitemdeclare{inproceedings}{el1999protocol}
\bibitem{el1999protocol}
\bibinfo{author}{Amal~El \surnamestart Fallah{-}Seghrouchni\surnameend},
  \bibinfo{author}{Serge \surnamestart Haddad\surnameend} \&
  \bibinfo{author}{Hamza \surnamestart Mazouzi\surnameend}
  (\bibinfo{year}{1999}): \emph{\bibinfo{title}{Protocol Engineering for
  Multi-agent Interaction}}.
\newblock In \bibinfo{editor}{Francisco~J. \surnamestart Garijo\surnameend} \&
  \bibinfo{editor}{Magnus \surnamestart Boman\surnameend}, editors: {\sl
  \bibinfo{booktitle}{MultiAgent System Engineering, 9th European Workshop on
  Modelling Autonomous Agents in a Multi-Agent World, {MAAMAW} '99, Valencia,
  Spain, June 30 - July 2, 1999, Proceedings}}, {\sl \bibinfo{series}{Lecture
  Notes in Computer Science}} \bibinfo{volume}{1647},
  \bibinfo{publisher}{Springer}, pp. \bibinfo{pages}{89--101},
  \doi{10.1007/3-540-48437-X\_8}.

\bibitemdeclare{inproceedings}{DBLP:conf/atal/FerrandoAM17}
\bibitem{DBLP:conf/atal/FerrandoAM17}
\bibinfo{author}{Angelo \surnamestart Ferrando\surnameend},
  \bibinfo{author}{Davide \surnamestart Ancona\surnameend} \&
  \bibinfo{author}{Viviana \surnamestart Mascardi\surnameend}
  (\bibinfo{year}{2017}): \emph{\bibinfo{title}{Decentralizing {MAS} Monitoring
  with DecAMon}}.
\newblock In \bibinfo{editor}{Kate \surnamestart Larson\surnameend},
  \bibinfo{editor}{Michael \surnamestart Winikoff\surnameend},
  \bibinfo{editor}{Sanmay \surnamestart Das\surnameend} \&
  \bibinfo{editor}{Edmund~H. \surnamestart Durfee\surnameend}, editors: {\sl
  \bibinfo{booktitle}{Proceedings of the 16th Conference on Autonomous Agents
  and MultiAgent Systems, {AAMAS} 2017, S{\~{a}}o Paulo, Brazil, May 8-12,
  2017}}, \bibinfo{publisher}{{ACM}}, pp. \bibinfo{pages}{239--248}.
\newblock \urlprefix\url{http://dl.acm.org/citation.cfm?id=3091164}.

\bibitemdeclare{inproceedings}{DBLP:conf/amcis/Fiedler19}
\bibitem{DBLP:conf/amcis/Fiedler19}
\bibinfo{author}{Kirk \surnamestart Fiedler\surnameend} (\bibinfo{year}{2019}):
  \emph{\bibinfo{title}{Virtual Reality in the Cloud: Amazon Sumerian as a Tool
  and Topic}}.
\newblock In: {\sl \bibinfo{booktitle}{25th Americas Conference on Information
  Systems, {AMCIS} 2019, Canc{\'{u}}n, Mexico, August 15-17, 2019}},
  \bibinfo{publisher}{Association for Information Systems}.
\newblock \urlprefix\url{https://aisel.aisnet.org/amcis2019/treo/treos/12}.

\bibitemdeclare{article}{DBLP:journals/corr/abs-2001-09124}
\bibitem{DBLP:journals/corr/abs-2001-09124}
\bibinfo{author}{Michael \surnamestart Fisher\surnameend},
  \bibinfo{author}{Viviana \surnamestart Mascardi\surnameend},
  \bibinfo{author}{Kristin~Yvonne \surnamestart Rozier\surnameend},
  \bibinfo{author}{Bernd{-}Holger \surnamestart Schlingloff\surnameend},
  \bibinfo{author}{Michael \surnamestart Winikoff\surnameend} \&
  \bibinfo{author}{Neil \surnamestart Yorke{-}Smith\surnameend}
  (\bibinfo{year}{2020}): \emph{\bibinfo{title}{Towards a Framework for
  Certification of Reliable Autonomous Systems}}.
\newblock {\sl \bibinfo{journal}{CoRR}} \bibinfo{volume}{abs/2001.09124}.
\newblock \urlprefix\url{https://arxiv.org/abs/2001.09124}.

\bibitemdeclare{article}{foster2019natural}
\bibitem{foster2019natural}
\bibinfo{author}{Mary~Ellen \surnamestart Foster\surnameend}
  (\bibinfo{year}{2019}): \emph{\bibinfo{title}{Natural language generation for
  social robotics: opportunities and challenges}}.
\newblock {\sl \bibinfo{journal}{Philosophical Transactions of the Royal
  Society B}} \bibinfo{volume}{374}(\bibinfo{number}{1771}), p.
  \bibinfo{pages}{20180027}, \doi{10.1098/rstb.2018.0027}.

\bibitemdeclare{article}{freund1999projective}
\bibitem{freund1999projective}
\bibinfo{author}{Eckhard \surnamestart Freund\surnameend} \&
  \bibinfo{author}{J{\"{u}}rgen \surnamestart Rossmann\surnameend}
  (\bibinfo{year}{1999}): \emph{\bibinfo{title}{Projective virtual reality:
  bridging the gap between virtual reality and robotics}}.
\newblock {\sl \bibinfo{journal}{{IEEE} Trans. Robotics Autom.}}
  \bibinfo{volume}{15}(\bibinfo{number}{3}), pp. \bibinfo{pages}{411--422},
  \doi{10.1109/70.768175}.

\bibitemdeclare{inproceedings}{gruber2008ontologies}
\bibitem{gruber2008ontologies}
\bibinfo{author}{Tom \surnamestart Gruber\surnameend} (\bibinfo{year}{2009}):
  \emph{\bibinfo{title}{Definition of Ontology}}.
\newblock In \bibinfo{editor}{Ling \surnamestart Liu\surnameend} \&
  \bibinfo{editor}{M.~Tamer \surnamestart \"{O}zsu\surnameend}, editors: {\sl
  \bibinfo{booktitle}{Encyclopedia of Database Systems}},
  \bibinfo{publisher}{Springer-Verlag}, pp. \bibinfo{pages}{1959--1959}.
\newblock
  \urlprefix\url{https://tomgruber.org/writing/ontology-definition-2007.htm},
  \doi{10.1016/0004-3702(80)90011-9}.

\bibitemdeclare{inproceedings}{guadarrama2013grounding}
\bibitem{guadarrama2013grounding}
\bibinfo{author}{Sergio \surnamestart Guadarrama\surnameend},
  \bibinfo{author}{Lorenzo \surnamestart Riano\surnameend},
  \bibinfo{author}{Dave \surnamestart Golland\surnameend},
  \bibinfo{author}{Daniel \surnamestart G{\"{o}}hring\surnameend},
  \bibinfo{author}{Yangqing \surnamestart Jia\surnameend}, \bibinfo{author}{Dan
  \surnamestart Klein\surnameend}, \bibinfo{author}{Pieter \surnamestart
  Abbeel\surnameend} \& \bibinfo{author}{Trevor \surnamestart
  Darrell\surnameend} (\bibinfo{year}{2013}): \emph{\bibinfo{title}{Grounding
  spatial relations for human-robot interaction}}.
\newblock In: {\sl \bibinfo{booktitle}{2013 {IEEE/RSJ} International Conference
  on Intelligent Robots and Systems, Tokyo, Japan, November 3-7, 2013}},
  \bibinfo{publisher}{{IEEE}}, pp. \bibinfo{pages}{1640--1647},
  \doi{10.1109/IROS.2013.6696569}.

\bibitemdeclare{inproceedings}{10.1145/3359996.3364245}
\bibitem{10.1145/3359996.3364245}
\bibinfo{author}{Florian \surnamestart Heinrich\surnameend},
  \bibinfo{author}{Kai \surnamestart Bornemann\surnameend},
  \bibinfo{author}{Kai \surnamestart Lawonn\surnameend} \&
  \bibinfo{author}{Christian \surnamestart Hansen\surnameend}
  (\bibinfo{year}{2019}): \emph{\bibinfo{title}{Depth Perception in Projective
  Augmented Reality: An Evaluation of Advanced Visualization Techniques}}.
\newblock In: {\sl \bibinfo{booktitle}{25th ACM Symposium on Virtual Reality
  Software and Technology}}, \bibinfo{series}{VRST ’19},
  \bibinfo{publisher}{Association for Computing Machinery},
  \bibinfo{address}{New York, NY, USA}, \doi{10.1145/3359996.3364245}.

\bibitemdeclare{inproceedings}{DBLP:conf/conversations/HeyselaarB19}
\bibitem{DBLP:conf/conversations/HeyselaarB19}
\bibinfo{author}{Evelien \surnamestart Heyselaar\surnameend} \&
  \bibinfo{author}{Tibor \surnamestart Bosse\surnameend}
  (\bibinfo{year}{2019}): \emph{\bibinfo{title}{Using Theory of Mind to Assess
  Users' Sense of Agency in Social Chatbots}}.
\newblock In \bibinfo{editor}{Asbj{\o}rn \surnamestart F{\o}lstad\surnameend},
  \bibinfo{editor}{Theo~B. \surnamestart Araujo\surnameend},
  \bibinfo{editor}{Symeon \surnamestart Papadopoulos\surnameend},
  \bibinfo{editor}{Effie~Lai{-}Chong \surnamestart Law\surnameend},
  \bibinfo{editor}{Ole{-}Christoffer \surnamestart Granmo\surnameend},
  \bibinfo{editor}{Ewa \surnamestart Luger\surnameend} \&
  \bibinfo{editor}{Petter~Bae \surnamestart Brandtz{\ae}g\surnameend}, editors:
  {\sl \bibinfo{booktitle}{Chatbot Research and Design - Third International
  Workshop, {CONVERSATIONS} 2019, Amsterdam, The Netherlands, November 19-20,
  2019, Revised Selected Papers}}, {\sl \bibinfo{series}{Lecture Notes in
  Computer Science}} \bibinfo{volume}{11970}, \bibinfo{publisher}{Springer},
  pp. \bibinfo{pages}{158--169}, \doi{10.1007/978-3-030-39540-7\_11}.

\bibitemdeclare{misc}{IEEE-P7007-standard}
\bibitem{IEEE-P7007-standard}
\bibinfo{author}{\surnamestart {Institute of Electrical and Electronics
  Engineers}\surnameend} (\bibinfo{year}{2017}): \emph{\bibinfo{title}{{P7007}
  -- Ontological Standard for Ethically Driven Robotics and Automation
  Systems}}.
\newblock \urlprefix\url{https://standards.ieee.org/project/7007.html}.

\bibitemdeclare{article}{Klinghammer2016AllocentricII}
\bibitem{Klinghammer2016AllocentricII}
\bibinfo{author}{Mathias \surnamestart Klinghammer\surnameend},
  \bibinfo{author}{Immo \surnamestart Sch{\"u}tz\surnameend},
  \bibinfo{author}{Gunnar \surnamestart Blohm\surnameend} \&
  \bibinfo{author}{Katja \surnamestart Fiehler\surnameend}
  (\bibinfo{year}{2016}): \emph{\bibinfo{title}{Allocentric information is used
  for memory-guided reaching in depth: A virtual reality study}}.
\newblock {\sl \bibinfo{journal}{Vision Research}} \bibinfo{volume}{129}, pp.
  \bibinfo{pages}{13--24},\doi{10.1016/j.visres.2016.10.004}.

\bibitemdeclare{inproceedings}{DBLP:conf/insci/KlopfensteinDR18}
\bibitem{DBLP:conf/insci/KlopfensteinDR18}
\bibinfo{author}{Lorenz~Cuno \surnamestart Klopfenstein\surnameend},
  \bibinfo{author}{Saverio \surnamestart Delpriori\surnameend} \&
  \bibinfo{author}{Alessio \surnamestart Ricci\surnameend}
  (\bibinfo{year}{2018}): \emph{\bibinfo{title}{Adapting a Conversational Text
  Generator for Online Chatbot Messaging}}.
\newblock In \bibinfo{editor}{Svetlana~S. \surnamestart Bodrunova\surnameend},
  \bibinfo{editor}{Olessia \surnamestart Koltsova\surnameend},
  \bibinfo{editor}{Asbj{\o}rn \surnamestart F{\o}lstad\surnameend},
  \bibinfo{editor}{Harry \surnamestart Halpin\surnameend},
  \bibinfo{editor}{Polina \surnamestart Kolozaridi\surnameend},
  \bibinfo{editor}{Leonid \surnamestart Yuldashev\surnameend},
  \bibinfo{editor}{Anna~S. \surnamestart Smoliarova\surnameend} \&
  \bibinfo{editor}{Heiko \surnamestart Niedermayer\surnameend}, editors: {\sl
  \bibinfo{booktitle}{Internet Science - {INSCI} 2018 International Workshops,
  St. Petersburg, Russia, October 24-26, 2018, Revised Selected Papers}}, {\sl
  \bibinfo{series}{Lecture Notes in Computer Science}} \bibinfo{volume}{11551},
  \bibinfo{publisher}{Springer}, pp. \bibinfo{pages}{87--99},
  \doi{10.1007/978-3-030-17705-8\_8}.

\bibitemdeclare{inproceedings}{kollar2014grounding}
\bibitem{kollar2014grounding}
\bibinfo{author}{Thomas \surnamestart Kollar\surnameend},
  \bibinfo{author}{Stefanie \surnamestart Tellex\surnameend},
  \bibinfo{author}{Deb \surnamestart Roy\surnameend} \&
  \bibinfo{author}{Nicholas \surnamestart Roy\surnameend}
  (\bibinfo{year}{2010}): \emph{\bibinfo{title}{Grounding Verbs of Motion in
  Natural Language Commands to Robots}}.
\newblock In \bibinfo{editor}{Oussama \surnamestart Khatib\surnameend},
  \bibinfo{editor}{Vijay \surnamestart Kumar\surnameend} \&
  \bibinfo{editor}{Gaurav~S. \surnamestart Sukhatme\surnameend}, editors: {\sl
  \bibinfo{booktitle}{Experimental Robotics - The 12th International Symposium
  on Experimental Robotics, {ISER} 2010, December 18-21, 2010, New Delhi and
  Agra, India}}, {\sl \bibinfo{series}{Springer Tracts in Advanced
  Robotics}}~\bibinfo{volume}{79}, \bibinfo{publisher}{Springer}, pp.
  \bibinfo{pages}{31--47}, \doi{10.1007/978-3-642-28572-1\_3}.

\bibitemdeclare{inproceedings}{DBLP:conf/ismar/KruijffSF10}
\bibitem{DBLP:conf/ismar/KruijffSF10}
\bibinfo{author}{Ernst \surnamestart Kruijff\surnameend},
  \bibinfo{author}{J.~Edward~Swan \surnamestart II\surnameend} \&
  \bibinfo{author}{Steven \surnamestart Feiner\surnameend}
  (\bibinfo{year}{2010}): \emph{\bibinfo{title}{Perceptual issues in augmented
  reality revisited}}.
\newblock In: {\sl \bibinfo{booktitle}{9th {IEEE} International Symposium on
  Mixed and Augmented Reality, {ISMAR} 2010, Seoul, Korea, 13-16 October
  2010}}, \bibinfo{publisher}{{IEEE} Computer Society}, pp.
  \bibinfo{pages}{3--12}, \doi{10.1109/ISMAR.2010.5643530}.

\bibitemdeclare{inproceedings}{DBLP:conf/ecis/LaumerMG19}
\bibitem{DBLP:conf/ecis/LaumerMG19}
\bibinfo{author}{Sven \surnamestart Laumer\surnameend},
  \bibinfo{author}{Christian \surnamestart Maier\surnameend} \&
  \bibinfo{author}{Fabian~Tobias \surnamestart Gubler\surnameend}
  (\bibinfo{year}{2019}): \emph{\bibinfo{title}{Chatbot Acceptance in
  Healthcare: Explaining User Adoption of Conversational Agents for disease
  Diagnosis}}.
\newblock In \bibinfo{editor}{Jan \surnamestart vom Brocke\surnameend},
  \bibinfo{editor}{Shirley \surnamestart Gregor\surnameend} \&
  \bibinfo{editor}{Oliver \surnamestart M{\"{u}}ller\surnameend}, editors: {\sl
  \bibinfo{booktitle}{27th European Conference on Information Systems -
  Information Systems for a Sharing Society, {ECIS} 2019, Stockholm and
  Uppsala, Sweden, June 8-14, 2019}}.

\bibitemdeclare{inproceedings}{DBLP:conf/icwe/LeottaAFORR18}
\bibitem{DBLP:conf/icwe/LeottaAFORR18}
\bibinfo{author}{Maurizio \surnamestart Leotta\surnameend},
  \bibinfo{author}{Davide \surnamestart Ancona\surnameend},
  \bibinfo{author}{Luca \surnamestart Franceschini\surnameend},
  \bibinfo{author}{Dario \surnamestart Olianas\surnameend},
  \bibinfo{author}{Marina \surnamestart Ribaudo\surnameend} \&
  \bibinfo{author}{Filippo \surnamestart Ricca\surnameend}
  (\bibinfo{year}{2018}): \emph{\bibinfo{title}{Towards a Runtime Verification
  Approach for Internet of Things Systems}}.
\newblock In \bibinfo{editor}{Cesare \surnamestart Pautasso\surnameend},
  \bibinfo{editor}{Fernando \surnamestart S{\'{a}}nchez{-}Figueroa\surnameend},
  \bibinfo{editor}{Kari \surnamestart Syst{\"{a}}\surnameend} \&
  \bibinfo{editor}{Juan Manuel~Murillo \surnamestart Rodriguez\surnameend},
  editors: {\sl \bibinfo{booktitle}{Current Trends in Web Engineering - {ICWE}
  2018 International Workshops, MATWEP, EnWot, KD-WEB, WEOD, TourismKG,
  C{\'{a}}ceres, Spain, June 5, 2018, Revised Selected Papers}}, {\sl
  \bibinfo{series}{Lecture Notes in Computer Science}} \bibinfo{volume}{11153},
  \bibinfo{publisher}{Springer}, pp. \bibinfo{pages}{83--96},
  \doi{10.1007/978-3-030-03056-8\_8}.

\bibitemdeclare{inproceedings}{DBLP:conf/enase/LeottaCFOARR19}
\bibitem{DBLP:conf/enase/LeottaCFOARR19}
\bibinfo{author}{Maurizio \surnamestart Leotta\surnameend},
  \bibinfo{author}{Diego \surnamestart Clerissi\surnameend},
  \bibinfo{author}{Luca \surnamestart Franceschini\surnameend},
  \bibinfo{author}{Dario \surnamestart Olianas\surnameend},
  \bibinfo{author}{Davide \surnamestart Ancona\surnameend},
  \bibinfo{author}{Filippo \surnamestart Ricca\surnameend} \&
  \bibinfo{author}{Marina \surnamestart Ribaudo\surnameend}
  (\bibinfo{year}{2019}): \emph{\bibinfo{title}{Comparing Testing and Runtime
  Verification of IoT Systems: {A} Preliminary Evaluation based on a Case
  Study}}.
\newblock In \bibinfo{editor}{Ernesto \surnamestart Damiani\surnameend},
  \bibinfo{editor}{George \surnamestart Spanoudakis\surnameend} \&
  \bibinfo{editor}{Leszek~A. \surnamestart Maciaszek\surnameend}, editors: {\sl
  \bibinfo{booktitle}{Proceedings of the 14th International Conference on
  Evaluation of Novel Approaches to Software Engineering, {ENASE} 2019,
  Heraklion, Crete, Greece, May 4-5, 2019}}, \bibinfo{publisher}{SciTePress},
  pp. \bibinfo{pages}{434--441}, \doi{10.5220/0007745604340441}.

\bibitemdeclare{article}{leue2014theoretical}
\bibitem{leue2014theoretical}
\bibinfo{author}{M~\surnamestart Leue\surnameend},
  \bibinfo{author}{TH~\surnamestart Jung\surnameend} et~al.
  (\bibinfo{year}{2014}): \emph{\bibinfo{title}{A theoretical model of
  augmented reality acceptance}}.
\newblock {\sl \bibinfo{journal}{E-review of Tourism Research}}
  \bibinfo{volume}{5}, \doi{10.1080/13683500.2015.1070801}.

\bibitemdeclare{inproceedings}{LVR19}
\bibitem{LVR19}
\bibinfo{author}{Jianwen \surnamestart Li\surnameend}, \bibinfo{author}{Moshe
  \surnamestart Vardi\surnameend} \& \bibinfo{author}{Kristin~Yvonne
  \surnamestart Rozier\surnameend} (\bibinfo{year}{2019}):
  \emph{\bibinfo{title}{Satisfiability Checking for Mission-Time {LTL}}}.
\newblock In: {\sl \bibinfo{booktitle}{Proceedings of 31st International
  Conference on Computer Aided Verification ({CAV}'19)}},
  \bibinfo{series}{LNCS}, \bibinfo{publisher}{Springer},
  \doi{10.1007/978-3-030-25543-5_1}.

\bibitemdeclare{inproceedings}{10.1145/3009939.3009947}
\bibitem{10.1145/3009939.3009947}
\bibinfo{author}{Martin \surnamestart Luboschik\surnameend},
  \bibinfo{author}{Philip \surnamestart Berger\surnameend} \&
  \bibinfo{author}{Oliver \surnamestart Staadt\surnameend}
  (\bibinfo{year}{2016}): \emph{\bibinfo{title}{On Spatial Perception Issues In
  Augmented Reality Based Immersive Analytics}}.
\newblock In: {\sl \bibinfo{booktitle}{Proceedings of the 2016 ACM Companion on
  Interactive Surfaces and Spaces}}, \bibinfo{series}{ISS ’16 Companion},
  \bibinfo{publisher}{Association for Computing Machinery},
  \bibinfo{address}{New York, NY, USA}, p. \bibinfo{pages}{47–53},
  \doi{10.1145/3009939.3009947}.

\bibitemdeclare{article}{maiello2015effectiveness}
\bibitem{maiello2015effectiveness}
\bibinfo{author}{Guido \surnamestart Maiello\surnameend},
  \bibinfo{author}{Manuela \surnamestart Chessa\surnameend},
  \bibinfo{author}{Fabio \surnamestart Solari\surnameend} \&
  \bibinfo{author}{Peter~J \surnamestart Bex\surnameend}
  (\bibinfo{year}{2015}): \emph{\bibinfo{title}{The (in) effectiveness of
  simulated blur for depth perception in naturalistic images}}.
\newblock {\sl \bibinfo{journal}{PloS one}}
  \bibinfo{volume}{10}(\bibinfo{number}{10}),\doi{10.1371/journal.pone.0140230.s004}.

\bibitemdeclare{article}{arr2020}
\bibitem{arr2020}
\bibinfo{author}{Zhanat \surnamestart Makhataeva\surnameend} \&
  \bibinfo{author}{Huseyin~Atakan \surnamestart Varol\surnameend}
  (\bibinfo{year}{2020}): \emph{\bibinfo{title}{Augmented Reality for Robotics:
  A Review}}.
\newblock {\sl \bibinfo{journal}{Robotics}}
  \bibinfo{volume}{9}(\bibinfo{number}{2}), \doi{10.3390/robotics9020021}.

\bibitemdeclare{inproceedings}{8864580}
\bibitem{8864580}
\bibinfo{author}{Sebastian \surnamestart von Mammen\surnameend},
  \bibinfo{author}{Andreas \surnamestart M{\"{u}}ller\surnameend},
  \bibinfo{author}{Marc~Erich \surnamestart Latoschik\surnameend},
  \bibinfo{author}{Mario \surnamestart Botsch\surnameend},
  \bibinfo{author}{Kirsten \surnamestart Brukamp\surnameend},
  \bibinfo{author}{Carsten \surnamestart Schr{\"{o}}der\surnameend} \&
  \bibinfo{author}{Michel \surnamestart Wacker\surnameend}
  (\bibinfo{year}{2019}): \emph{\bibinfo{title}{{VIA} {VR:} {A} Technology
  Platform for Virtual Adventures for Healthcare and Well-Being}}.
\newblock In \bibinfo{editor}{Fotis \surnamestart Liarokapis\surnameend},
  editor: {\sl \bibinfo{booktitle}{11th International Conference on Virtual
  Worlds and Games for Serious Applications, VS-Games 2019, Vienna, Austria,
  September 4-6, 2019}}, \bibinfo{publisher}{{IEEE}}, pp.
  \bibinfo{pages}{1--2}, \doi{10.1109/VS-Games.2019.8864580}.

\bibitemdeclare{book}{maynard2016natural}
\bibitem{maynard2016natural}
\bibinfo{author}{Diana \surnamestart Maynard\surnameend},
  \bibinfo{author}{Kalina \surnamestart Bontcheva\surnameend} \&
  \bibinfo{author}{Isabelle \surnamestart Augenstein\surnameend}
  (\bibinfo{year}{2016}): \emph{\bibinfo{title}{Natural Language Processing for
  the Semantic Web}}.
\newblock \bibinfo{series}{Synthesis Lectures on the Semantic Web: Theory and
  Technology}, \bibinfo{publisher}{Morgan {\&} Claypool Publishers},
  \doi{10.2200/S00741ED1V01Y201611WBE015}.

\bibitemdeclare{article}{mellish2005semantic}
\bibitem{mellish2005semantic}
\bibinfo{author}{Chris \surnamestart Mellish\surnameend} \&
  \bibinfo{author}{Xiantang \surnamestart Sun\surnameend}
  (\bibinfo{year}{2006}): \emph{\bibinfo{title}{The semantic web as a
  Linguistic resource: Opportunities for natural language generation}}.
\newblock {\sl \bibinfo{journal}{Knowl. Based Syst.}}
  \bibinfo{volume}{19}(\bibinfo{number}{5}), pp. \bibinfo{pages}{298--303},
  \doi{10.1016/j.knosys.2005.11.011}.

\bibitemdeclare{unpublished}{interaction}
\bibitem{interaction}
\bibinfo{author}{\surnamestart {Merriam-Webster Dictionary}\surnameend}
  (\bibinfo{year}{2020}): \emph{\bibinfo{title}{Definition of Interaction}}.
\newblock
  \urlprefix\url{https://www.merriam-webster.com/dictionary/interaction}.

\bibitemdeclare{inproceedings}{Milgram94augmentedreality}
\bibitem{Milgram94augmentedreality}
\bibinfo{author}{Paul \surnamestart Milgram\surnameend}, \bibinfo{author}{Haruo
  \surnamestart Takemura\surnameend}, \bibinfo{author}{Akira \surnamestart
  Utsumi\surnameend} \& \bibinfo{author}{Fumio \surnamestart
  Kishino\surnameend} (\bibinfo{year}{1994}): \emph{\bibinfo{title}{Augmented
  Reality: A Class of Displays on the Reality-Virtuality Continuum}}.
\newblock In: {\sl \bibinfo{booktitle}{Proceedings of Telemanipulator and
  Telepresence Technologies}}, \bibinfo{volume}{2351},
  \bibinfo{publisher}{SPIE}, pp. \bibinfo{pages}{282--292},
  \doi{10.1117/12.197321}.

\bibitemdeclare{inproceedings}{DBLP:conf/amcis/MutterleinH17}
\bibitem{DBLP:conf/amcis/MutterleinH17}
\bibinfo{author}{Joschka \surnamestart M{\"{u}}tterlein\surnameend} \&
  \bibinfo{author}{Thomas \surnamestart Hess\surnameend}
  (\bibinfo{year}{2017}): \emph{\bibinfo{title}{Immersion, Presence,
  Interactivity: Towards a Joint Understanding of Factors Influencing Virtual
  Reality Acceptance and Use}}.
\newblock In: {\sl \bibinfo{booktitle}{23rd Americas Conference on Information
  Systems, {AMCIS} 2017, Boston, MA, USA, August 10-12, 2017}},
  \bibinfo{publisher}{Association for Information Systems}.
\newblock
  \urlprefix\url{http://aisel.aisnet.org/amcis2017/AdoptionIT/Presentations/17}.

\bibitemdeclare{inproceedings}{odell2000representing}
\bibitem{odell2000representing}
\bibinfo{author}{James \surnamestart Odell\surnameend},
  \bibinfo{author}{H.~Van~Dyke \surnamestart Parunak\surnameend} \&
  \bibinfo{author}{Bernhard \surnamestart Bauer\surnameend}
  (\bibinfo{year}{2000}): \emph{\bibinfo{title}{Representing Agent Interaction
  Protocols in {UML}}}.
\newblock In \bibinfo{editor}{Paolo \surnamestart Ciancarini\surnameend} \&
  \bibinfo{editor}{Michael~J. \surnamestart Wooldridge\surnameend}, editors:
  {\sl \bibinfo{booktitle}{Agent-Oriented Software Engineering, First
  International Workshop, {AOSE} 2000, Limerick, Ireland, June 10, 2000,
  Revised Papers}}, {\sl \bibinfo{series}{Lecture Notes in Computer Science}}
  \bibinfo{volume}{1957}, \bibinfo{publisher}{Springer}, pp.
  \bibinfo{pages}{121--140}, \doi{10.1007/3-540-44564-1\_8}.

\bibitemdeclare{inproceedings}{pointon2018affordances}
\bibitem{pointon2018affordances}
\bibinfo{author}{Grant \surnamestart Pointon\surnameend},
  \bibinfo{author}{Chelsey \surnamestart Thompson\surnameend},
  \bibinfo{author}{Sarah \surnamestart Creem-Regehr\surnameend},
  \bibinfo{author}{Jeanine \surnamestart Stefanucci\surnameend} \&
  \bibinfo{author}{Bobby \surnamestart Bodenheimer\surnameend}
  (\bibinfo{year}{2018}): \emph{\bibinfo{title}{Affordances as a measure of
  perceptual fidelity in augmented reality}}.
\newblock In: {\sl \bibinfo{booktitle}{Proceedings of the 2018 IEEE VR 2018
  Workshop on Perceptual and Cognitive Issues in AR (PERCAR)}}, pp.
  \bibinfo{pages}{1--6}.

\bibitemdeclare{inproceedings}{rietz2019impact}
\bibitem{rietz2019impact}
\bibinfo{author}{Tim \surnamestart Rietz\surnameend}, \bibinfo{author}{Ivo
  \surnamestart Benke\surnameend} \& \bibinfo{author}{Alexander \surnamestart
  Maedche\surnameend} (\bibinfo{year}{2019}): \emph{\bibinfo{title}{The Impact
  of Anthropomorphic and Functional Chatbot Design Features in Enterprise
  Collaboration Systems on User Acceptance}}.
\newblock In \bibinfo{editor}{Thomas \surnamestart Ludwig\surnameend} \&
  \bibinfo{editor}{Volkmar \surnamestart Pipek\surnameend}, editors: {\sl
  \bibinfo{booktitle}{Human Practice. Digital Ecologies. Our Future. 14.
  Internationale Tagung Wirtschaftsinformatik {(WI} 2019), February 24-27,
  2019, Siegen, Germany}}, \bibinfo{publisher}{University of Siegen, Germany /
  AISeL}, pp. \bibinfo{pages}{1642--1656}.
\newblock \urlprefix\url{https://aisel.aisnet.org/wi2019/track13/papers/7}.

\bibitemdeclare{article}{rossi2017user}
\bibitem{rossi2017user}
\bibinfo{author}{Silvia \surnamestart Rossi\surnameend},
  \bibinfo{author}{Fran{\c{c}}ois \surnamestart Ferland\surnameend} \&
  \bibinfo{author}{Adriana \surnamestart Tapus\surnameend}
  (\bibinfo{year}{2017}): \emph{\bibinfo{title}{User profiling and behavioral
  adaptation for {HRI:} {A} survey}}.
\newblock {\sl \bibinfo{journal}{Pattern Recognit. Lett.}}
  \bibinfo{volume}{99}, pp. \bibinfo{pages}{3--12},
  \doi{10.1016/j.patrec.2017.06.002}.

\bibitemdeclare{inproceedings}{10.1007}
\bibitem{10.1007}
\bibinfo{author}{Daniel \surnamestart Szafir\surnameend}
  (\bibinfo{year}{2019}): \emph{\bibinfo{title}{Mediating Human-Robot
  Interactions with Virtual, Augmented, and Mixed Reality}}.
\newblock In \bibinfo{editor}{Jessie Y.~C. \surnamestart Chen\surnameend} \&
  \bibinfo{editor}{Gino \surnamestart Fragomeni\surnameend}, editors: {\sl
  \bibinfo{booktitle}{Virtual, Augmented and Mixed Reality. Applications and
  Case Studies - 11th International Conference, {VAMR} 2019, Held as Part of
  the 21st {HCI} International Conference, {HCII} 2019, Orlando, FL, USA, July
  26-31, 2019, Proceedings, Part {II}}}, {\sl \bibinfo{series}{Lecture Notes in
  Computer Science}} \bibinfo{volume}{11575}, \bibinfo{publisher}{Springer},
  pp. \bibinfo{pages}{124--149}, \doi{10.1007/978-3-030-21565-1\_9}.

\bibitemdeclare{inproceedings}{DBLP:conf/semeval/TafreshiD19}
\bibitem{DBLP:conf/semeval/TafreshiD19}
\bibinfo{author}{Shabnam \surnamestart Tafreshi\surnameend} \&
  \bibinfo{author}{Mona~T. \surnamestart Diab\surnameend}
  (\bibinfo{year}{2019}): \emph{\bibinfo{title}{{GWU} {NLP} Lab at SemEval-2019
  Task 3 : EmoContext: Effectiveness ofContextual Information in Models for
  Emotion Detection inSentence-level at Multi-genre Corpus}}.
\newblock In \bibinfo{editor}{Jonathan \surnamestart May\surnameend},
  \bibinfo{editor}{Ekaterina \surnamestart Shutova\surnameend},
  \bibinfo{editor}{Aur{\'{e}}lie \surnamestart Herbelot\surnameend},
  \bibinfo{editor}{Xiaodan \surnamestart Zhu\surnameend},
  \bibinfo{editor}{Marianna \surnamestart Apidianaki\surnameend} \&
  \bibinfo{editor}{Saif~M. \surnamestart Mohammad\surnameend}, editors: {\sl
  \bibinfo{booktitle}{Proceedings of the 13th International Workshop on
  Semantic Evaluation, SemEval@NAACL-HLT 2019, Minneapolis, MN, USA, June 6-7,
  2019}}, \bibinfo{publisher}{Association for Computational Linguistics}, pp.
  \bibinfo{pages}{230--235}, \doi{10.18653/v1/s19-2038}.

\bibitemdeclare{book}{EthicallyAlignedDesign2019}
\bibitem{EthicallyAlignedDesign2019}
\bibinfo{editor}{\surnamestart {The IEEE Global Initiative on Ethics of
  Autonomous and Intelligent Systems}\surnameend}, editor
  (\bibinfo{year}{2019}): \emph{\bibinfo{title}{Ethically Aligned Design: A
  Vision for Prioritizing Human Well-being with Autonomous and Intelligent
  Systems}}.
\newblock \bibinfo{publisher}{IEEE}.
\newblock
  \urlprefix\url{https://standards.ieee.org/content/ieee-standards/en/industry-connections/ec/
  autonomous-systems.html}.

\bibitemdeclare{article}{uschold1996ontologies}
\bibitem{uschold1996ontologies}
\bibinfo{author}{Mike \surnamestart Uschold\surnameend} \&
  \bibinfo{author}{Michael \surnamestart Gruninger\surnameend}
  (\bibinfo{year}{1996}): \emph{\bibinfo{title}{Ontologies: Principles, methods
  and applications}}.
\newblock {\sl \bibinfo{journal}{The knowledge engineering review}}
  \bibinfo{volume}{11}(\bibinfo{number}{2}), pp. \bibinfo{pages}{93--136},
  \doi{10.1017/S0269888900007797}.

\bibitemdeclare{article}{wilks2009natural}
\bibitem{wilks2009natural}
\bibinfo{author}{Yorick \surnamestart Wilks\surnameend} \&
  \bibinfo{author}{Christopher \surnamestart Brewster\surnameend}
  (\bibinfo{year}{2009}): \emph{\bibinfo{title}{Natural Language Processing as
  a Foundation of the Semantic Web}}.
\newblock {\sl \bibinfo{journal}{Found. Trends Web Sci.}}
  \bibinfo{volume}{1}(\bibinfo{number}{3-4}), pp. \bibinfo{pages}{199--327},
  \doi{10.1561/1800000002}.

\bibitemdeclare{article}{DBLP:journals/pieee/WinfieldMPE19}
\bibitem{DBLP:journals/pieee/WinfieldMPE19}
\bibinfo{author}{Alan F.~T. \surnamestart Winfield\surnameend},
  \bibinfo{author}{Katina \surnamestart Michael\surnameend},
  \bibinfo{author}{Jeremy \surnamestart Pitt\surnameend} \&
  \bibinfo{author}{Vanessa \surnamestart Evers\surnameend}
  (\bibinfo{year}{2019}): \emph{\bibinfo{title}{Machine Ethics: The Design and
  Governance of Ethical {AI} and Autonomous Systems}}.
\newblock {\sl \bibinfo{journal}{Proceedings of the {IEEE}}}
  \bibinfo{volume}{107}(\bibinfo{number}{3}), pp. \bibinfo{pages}{509--517},
  \doi{10.1109/JPROC.2019.2900622}.

\end{thebibliography}

\end{document}